\documentclass[a4paper, 10pt, oneside]{article}
\usepackage{jheppub}
\usepackage{amssymb}
\usepackage{amsmath,graphicx,color,epsfig,graphics}
\usepackage{subfig}
\usepackage{pstricks}
\usepackage{float}
\usepackage{slashed}
\newcommand\fverb{\setbox\fverbbox=\hbox\bgroup\verb}
\newcommand\fverbdo{\egroup\medskip\noindent%
\fbox{\unhbox\fverbbox}\ }
\newcommand\fverbit{\egroup\item[\fbox{\unhbox\fverbbox}]}
\newbox\fverbbox

\begin{document}
\title{Lepton Polarization Asymmetries of $H\to\gamma\tau^+\tau^-$ Decays in Standard Model}
\author[a,b]{Rabia Akbar,}
\author[a,b]{Ishtiaq Ahmed,}
\author[b]{M. Jamil Aslam}
\affiliation[a]{National Centre for Physics,\\
Quaid-i-Azam University Campus, Islamabad 45320, Pakistan}
\affiliation[b]{Department of Physics,\\
Quaid-i-Azam University, Islamabad 45320, Pakistan}
\emailAdd{rabia.akbar@ncp.edu.pk}\emailAdd{ishtiaq@ncp.edu.pk}
\emailAdd{jamil@ncp.edu.pk}
\date{\today}

\abstract{Recently, CMS and ATLAS collaborations at LHC announced
a Higgs like particle with mass near $125$GeV. Regarding this, to
explore its intrinsic properties, different observables are
needed to be measured precisely at the LHC for various decay channels of the
Higgs. In this context, we calculate the final state lepton
polarization asymmetries, namely, single lepton polarization asymmetries ($P_i$) and
double lepton polarization asymmetries ($P_{ij}$) in the SM for
radiative semileptonic Higgs decay $H\to\gamma\tau^+\tau^-$.
In the phenomenological analysis of
these lepton polarization asymmetries both tree and loop level diagrams are considered and
it is found that these diagrams give important contributions
in the evaluation of said asymmetries. Interestingly, it is
found that in $P_{ij}$ the tree level diagrams contribute
separately, which however, are missing in the calculations of $P_i$ and the lepton forward-backward
asymmetries ($A_{FB}$). Similar to the other observables such as
the decay rate and the lepton forward-backward asymmetries, the $\tau$-lepton
polarization asymmetries would be interesting observables.
The experimental study of these
observables will provide a fertile ground to explore the intrinsic properties of the
SM Higgs boson and its dynamics as well as help us to extract the
signatures of the possible new physics beyond the SM.}

\maketitle

\section{Introduction}

In July 2012, the discovery of a new heavy Higgs like particle with a mass
around $125$GeV announced at CERN by the ATLAS and CMS experiments \cite{4,5}, led us
to an exciting era of the particle physics. This is a great leap towards the success of a
theory proposed by Glashow, Salam and Weinberg in 1970, where it was realized that there is a close ties
between the electromagnetic and the weak force and these are the manifestation of a single underlying force, the
electroweak force. The electroweak unification of the forces is presently known as the Standard Model (SM).
Since last four decades the SM has been tested by many experiments and it has been shown to successfully
describe high-energy particle interactions. The simplest and most elegant way to construct the SM would be to
assert that all fundamental particles are massless, and by the virtue of a Higgs mechanism the photon remains massless while its close cousins, the $W$ and $Z$ bosons, acquire a mass some $100$ times that of a proton mass. Experimentally these bosons were discovered with the mass
predicted in the SM and the discovery of the Higgs boson is a main missing chunk of this model.

With the advent of a Higgs like boson, the next goal of the particles physics is to understand its nature
to be sure if it is the Higgs boson of the SM or a new scalar particle. This can be done by studying the different possible
decays of the Higgs boson and it is an important task for the theoretical physics. A numerous amount of literature is already
present on the theoretical study
of the SM Higgs boson. On the other hand in experimental analysis of different decay channels of a newly
observed particle at the ATLAS \cite{4} and CMS \cite{5} experiments suggest
that it is most likely a signal of the SM Higgs boson. However, it
is still needed to be confirmed whether it is a SM Higgs boson or not?
because of the excess events of $H\to\gamma\gamma$ decay. In this
context, the diphoton decay channel of Higgs is studied in different new physics
scenarios \cite{N1,N1a,N1b,N1c,N2,N2a,N2b,N2c,N2d,N2e,N2f,N3,N3a,N3b,N3c,N3d,N3e,N3f,N3g,N31h,N3h,N3i,N3j,N3k,N3l,N3m},
but due to the limitation of data it will take sometime to distinguish these beyond SM signatures.

Furthermore, to have more insight about the Higgs boson
properties, besides the $H\to\gamma\gamma$ decay channel, a
complementary radiative decay channel $H\to\gamma\ell^+\ell^-$ with
$\ell=e,$ $\mu$ or $\tau$ got some attentions \cite{6, 8, 9, 10,
11, AFB1}. In these studies, the emphasis is on the analysis of the decay rates,
invariant mass distributions and leptons forward-backward asymmetry
$(A_{FB})$. This channel is actually induced by $H\to\gamma\gamma$
through the internal conversion i.e. the decay of a virtual photon
to a pair of leptons. For the numerical calculations of aforementioned observables
performed in refs. \cite{10, 11, AFB1} the value of the mass of the Higgs
particle is set to be 125GeV. Furthermore, Sun \textit{et. al.} \cite{AFB1} focused on the
calculation of the lepton forward-backward
asymmetries $(A_{FB})$ in $H\to\gamma\ell^+\ell^-$ decays. They have shown that due to the parity odd decay
of the Higgs $H\to \gamma Z^*$, the lepton forward-backward asymmetry is expected to be non zero in the semileptonic Higgs boson decays.
The values of the $A_{FB}$ in case of electrons and muons as final state leptons come out to be of the order of $10^{-2}$
which will be a challenging task to measure at the LHC. This motivated us to look for the other
asymmetries which may have larger magnitude than the $A_{FB}$. Going along this direction we performed the detailed study
of a single and the double lepton polarization asymmetries for $H\to\gamma\tau^+\tau^-$ decays by closely following the scheme of the study of
$A_{FB}$ performed in ref. \cite{AFB1}. In principle one can include the case where electrons and $\mu$'s appear as the final state leptons but it has some technical issues.

The first one is that at the LHC detectors (CMS and ATLAS)
there is a possibility to reconstruct the polarization of a particle if it is unstable because the spin direction is inferred from the decay
distribution \cite{Sug1, Sug2}. In this way the electrons are stable so their polarization can not be detected via energy measurement of final state. Also muons produced in the Higgs decay are penetrating and they can travel an average distance of 400$km$, so its reconstruction inside the LHC detectors
is impossible. The second one is due to the low event rates of the $H \to \gamma e^+ e^-$ and $H \to \gamma \mu^+ \mu^-$ decay channels. To elaborate this point, if one assume that the Higgs production is mainly driven by the gluon-gluon fusion, then even for LHC operating at its full energy, i.e. 14 TeV, the roughly estimates cross-sections are the
\begin{eqnarray}
\sigma(pp\to H \to \gamma e^+ e^-)&\thickapprox& 3 \times 10^{-2} fb \notag \\
\sigma(pp\to H \to \gamma \mu^+ \mu^-)&\thickapprox& 5 \times 10^{-2}fb \notag \\
\sigma(pp\to H \to \gamma \tau^+ \tau^-)&\thickapprox& 4.2 fb. \label{events}
\end{eqnarray}

The integrated luminosity of $300 fb^{-1}$ is expected from the upcoming run period, therefore, we can
expect $\mathcal{O}(10)$ events of $H \to \gamma e^+ e^-$ and $H \to \gamma \mu^+ \mu^-$ decays, even without considering the background
effects. These number of events are too few for any appreciable measurements of an observable such as the polarization asymmetry. Due to these reasons, the polarizations of electron and muon are not measured at the ATLAS and CMS detectors, therefore, we will not add their numerical analysis in the forthcoming study.

The situation for the $\tau$ lepton is different because at first it is a sequential lepton so its decay is maximally parity violating. In the $\tau$
rest frame, parity violation determines the angular distribution of the $\tau$ product with respect to $\tau$ helicity. When boosted into the
lab frame this angular distribution is manifest in the form of the energy and angular distribution of the decay products which can be measured. Thus the energy and angular distributions of $\tau$ decay products is a $\tau$ polarimeter. Also, the $\tau$ decay length at the centre of
mass energy of the Higgs mass $\sqrt{s} = M_{H}$ is around $3.6mm$ and this ensure
that $\tau$ decays are easily contained within the detector. Just as an example,
$\tau$ polarization has been measured by the ATLAS in $\tau$ hadronic decays with a
single final state charged particle \cite{Sug3}. In addition, compared to the electrons and muons as the final state leptons, the Higgs decaying to
taus will give a thousand of events (c.f. Eq. (\ref{events})) and hence it is likely that an analysis can be performed. Therefore, the precise
measurements of these asymmetries in future will not only increase our understanding of the properties of Higgs boson but will
also give us a valuable information about its various couplings.

The structure of the paper is as follows. In section 2, we present the
theoretical framework necessary for $H\to\gamma\tau^+\tau^-$ decays.
After defining the formulae of asymmetries under consideration in
section 3, we derive the expressions of these asymmetries. In
section 4, the above mentioned observables will be analyzed numerically and will be discussed in length. Finally, in the last
section we will summarize and conclude the main results.  Appendix
includes some definitions and the fermion and $W$-boson loops functions.
\begin{figure}
\centering
\includegraphics[scale=0.4]{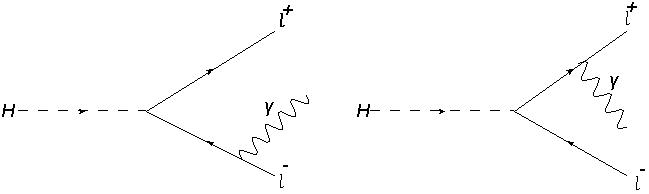}
\caption{The tree level diagrams for $H\to\gamma\ell^+\ell^-$
decays, where $\ell$ corresponds to $\tau$.}\label{treediag}
\end{figure}

\section{Formulation of the Amplitude}
The tree and the loop level Feynman diagrams for  $H\to\gamma\tau^+\tau^-$ decays
are shown in Figs. 1 and 2, respectively. The
amplitudes for these diagrams can be expressed as \cite{AFB1}
\begin{eqnarray}
\mathcal{M}_{tree}&=&\mathcal{C}_{0}
\bar{u}({p_2})\bigg(\frac{2p_{2}^{\nu}+\gamma^{\nu}\slashed k}{2p_{2}\dot k}-\frac{\slashed k \gamma^{\nu}+2p_{1}^{\nu}}{2p_{1} \dot k}
\bigg)v({p_1})\epsilon_{\nu}^{*} ,\notag \\
\mathcal{M}_{Loop}&=&\epsilon^{*\nu}(k_{\mu}q_{\nu}-g_{\mu\nu}(k.q))\bar{u}(p_2)(\mathcal{C}_{1}\gamma^{\mu}+\mathcal{C}_{2}\gamma^{\mu}\gamma^{5})v(p_{1})\notag\\
&&\quad+\epsilon_{\mu\nu\alpha\beta}\epsilon^{*\nu}
k^{\alpha}q^{\beta}\bar{u}(p_2)(\mathcal{C}_{3}\gamma^{\mu}+\mathcal{C}_{4}\gamma^{\mu}\gamma^{5})v(p_1),
\end{eqnarray}
\newline
\begin{figure}
\centering
\includegraphics[scale=0.4]{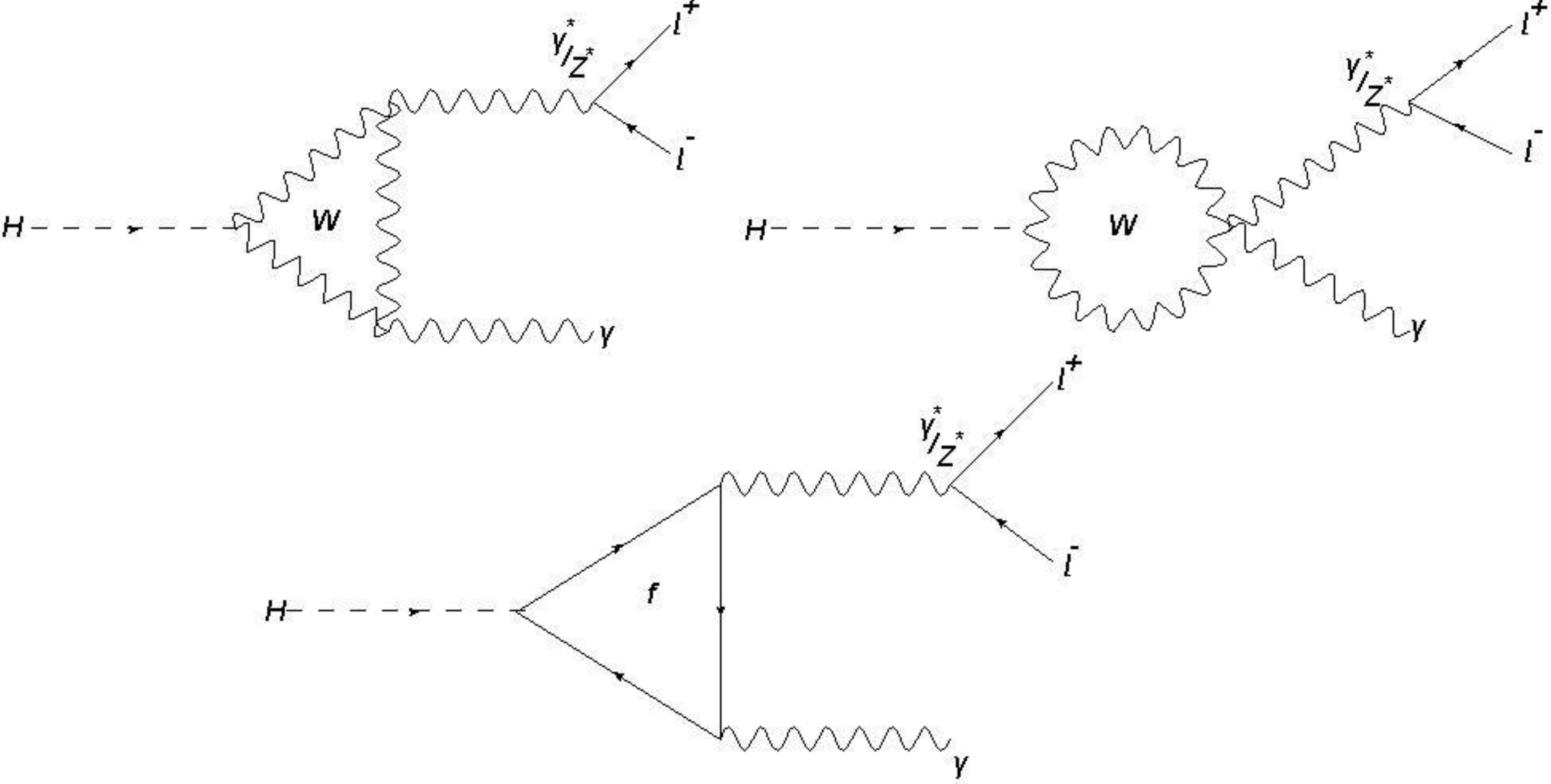}
\caption{The $Z^*$ and $\gamma^*$ loop diagrams for
$H\to\gamma\ell^+\ell^-$ decays, with $\ell$ corresponding to $\tau$ lepton.}\label{loopdiag}
\end{figure}
with
\begin{eqnarray}
\mathcal C_{0}&=& \frac{-2\pi \alpha
m_{\tau}}{m_{W}sin\theta_{W}},\notag \\
\mathcal C_{1}&=&-(\frac{1}{4}-sin^{2}\theta_{W})P_{Z}\Pi_{s\gamma
Z}-\frac{1}{s}\Pi_{\gamma \gamma},\notag \qquad \qquad
\mathcal C_{2}=\frac{1}{4}P_{Z}\Pi_{s\gamma Z},\notag \\
\mathcal C_{3}&=&-(\frac{1}{4}-sin^{2}\theta_{W})P_{Z}\Pi_{a\gamma
Z} \qquad \qquad\qquad\qquad \mathcal C_{4}=\frac{1}{4}P_{Z}\Pi_{a
\gamma Z}.\label{Cfunctions}
\end{eqnarray}
Here, $m_\tau$ denotes the mass of $\tau$ lepton, $p_{1}$, $p_{2}$, $k$ and
$q$ are the momenta of $\tau^{-}$, $\tau^{+}$, $\gamma $ and the virtual
particles $\gamma^*$ or $Z^*$, respectively. The fine structure constant is $\alpha$
while $\theta_W$ is the electroweak mixing angle.
The definitions of the $P_Z$, $\Pi_{s\gamma Z}$, $\Pi_{\gamma
\gamma}$ and $\Pi_{a \gamma Z}$ are given in ref. \cite{AFB1},
however, for the sake of completeness we have recollected them
in the Appendix A.

Using these amplitudes, we have derived the expression of the
the decay rate for the process $H \rightarrow \gamma
\tau^{+}\tau^{-}$. Except a negative sign in the
third term of the numerator of $\mathcal{B}$ defined below, our expression of the differential decay rate of $H \to \gamma \tau^{+}\tau^{-}$ decay is similar to the one given in ref. \cite{AFB1} and it can be
expressed as follow:
\begin{eqnarray}
\frac{d^2\Gamma}{dsdx}=\frac{(m_{H}^2-s)}{512\pi^3m_{H}^3}v\Delta,
\end{eqnarray}
with
\begin{eqnarray}
\Delta&=&|\mathcal C_{0}|^{2}
\mathcal{A}+2Re(\mathcal C_{0} C_{1}^*)\mathcal{B}+ 2 Re( \mathcal C_{0}\mathcal C_{4}^*)\mathcal{C}\notag\\
&&+(|\mathcal C_1|^2+|\mathcal C_{3}|^2)\mathcal{D}+(|\mathcal C_2|^2+|\mathcal C_{4}|^2)\mathcal{E}+2Re(\mathcal C_{1}\mathcal C_{4}^{*}+\mathcal C_{2}\mathcal C_{3}^*)\mathcal{F},\label{delta}
\end{eqnarray}
and
\begin{eqnarray}
\mathcal{A}&=&\frac{1}{(m_{H}^2-s)^2(1-v^2
x^2)^2}[m_{H}^4+s^2+32m_{\tau}^4-8m_{\tau}^{2}s-8m_{\tau}^2
m_{H}^2-(m_{H}^4+ s^2 -8m_{\tau}^2 s)v^2 x^2], \notag\\
\mathcal{B}&=&8m_{\tau}\frac{m_{H}^2-s-s v^{2}(1-x^{2})}{1-v^{2}
x^{2}},\notag\\
\mathcal{C}&=&8m_{\tau}\frac{(m_{H}^2-s)}{1-v^{2} x^{2}}
v x,\notag\\
\mathcal{D}&=&\frac{(m_{H}^2-s)^2}{2}(s +4 m_{\tau}^2+s ^{2}
v^2x^{2}),\notag\\
\mathcal{E}&=&\frac{(m_{H}^2-s)^2}{2} s v^2 (1+x^{2}),\notag\\
\mathcal{F}&=&(m_{H}^2-s)^2 s vx.\label{ABCfunctions}
\end{eqnarray}
Here, $s$ is the square of momentum transfer i.e., $s=q^2$, $v=\sqrt{1-\frac{4m_{\tau}^2}{s}}$ and $ x= cos\theta$ with $\theta$ is the angle between Higgs boson and a
lepton in the rest frame of dileptons. The limits on the phase space parameters $s$ and $x$ are
\begin{eqnarray}
4m_{\tau}^2 \leq s  \leq m_{H}^2,  \qquad   -1\leq x \leq +1.
\end{eqnarray}

In Eq. (\ref{ABCfunctions}), the term $\mathcal{A}$ represents the contribution from the tree diagrams where as
$\mathcal{B}$ and $\mathcal{C}$ terms correspond to the interference
between the tree and loop diagrams. The last three terms
$\mathcal{D}$, $\mathcal{E}$ and $\mathcal{F}$ are originated purely from
the loop diagrams.

\section{Observables}
It has already mentioned that the differential decay rate and the lepton forward-backward asymmetry
were under debate in literature \cite{6, 8, 9, 10, 11, AFB1} and the purpose of this study is to
investigate the polarization asymmetries of the final state $\tau$ leptons in $H \to \gamma \tau^{+}\tau^{-}$ decay.
To achieve this goal, let us introduce the orthogonal four vectors belonging to the
polarization of $\tau^-$ and $\tau^+$, namely $S_i^{-}$ and $S_i^{+}$, respectively \cite{Polarization1,Polarization2}. These
polarization vectors can be defined as follow:
\begin{eqnarray}
S_L^{-\alpha}\equiv(0,\textbf{e}_L)&=&\left(0,\frac{\textbf{p}_1}{|\textbf{p}_1|}\right),\notag\\
S_N^{-\alpha}\equiv(0,\textbf{e}_N)&=&\left(0,\frac{\textbf{k}\times\textbf{p}_1}{|\textbf{k}\times\textbf{p}_1|}\right),\notag\\
S_T^{-\alpha}\equiv(0,\textbf{e}_T)&=&(0,\textbf{e}_N\times\textbf{e}_L),\label{PV1}
\end{eqnarray}
where the subscripts $L$, $N$ and $T$ correspond to the longitudinal, normal and transverse polarizations, respectively.
Also,
$\textbf{p}_1$, $\textbf{p}_2$ and $\textbf{k}$ are denoting the three momenta vectors of the
final particles $\tau^-$, $\tau^+$ and $\gamma$, respectively, in the centre of mass frame of $\tau^{+}\tau^{-}$ system. It can be noticed that by replacing $\textbf{p}_1\to\textbf{p}_2$ one can
obtain the unit vectors for the polarizations of $\tau^+$. The longitudinal unit vector
$S_{L}$ are boosted by Lorentz transformations in the CM frame of $\tau^{+}\tau^{-}$:
\begin{eqnarray}
 S_{LCM}^{-\alpha}&=&\left(\frac{|\textbf{p}_1|}{m_{\tau}},\frac{E_{l}\textbf{p}_1}{m_{\tau}|\textbf{p}_1|}\right).\label{PV3}
\end{eqnarray}
With these unit vectors, let us define the single lepton polarization asymmetry in the following way:
\begin{eqnarray}
P_{i}^{(\pm)}(s) &=&\frac{\frac{d\Gamma}{ds}(\textbf{n}^\pm = \textbf{e}_{i}^{\pm})-\frac{d\Gamma}{ds}(\textbf{n}^\pm = -\textbf{e}_{i}^{\pm})}{\frac{d\Gamma}{ds}(\textbf{n}^\pm = \textbf{e}_{i}^{\pm})+\frac{d\Gamma}{ds}(\textbf{n}^\pm = -\textbf{e}_{i}^{\pm})},\label{singlepol}
\end{eqnarray}
where $\textbf{e}_{i}$ denotes the unit vector with subscript $i$ corresponds to longitudinal $(L)$, normal $(N)$ and transverse $(T)$ lepton polarizations
index, and $\textbf{n}^\pm$ is the spin direction of $\tau^\pm$. The differential decay rate for the polarized lepton $\tau^{\pm}$
in $H\to \gamma \tau^{+}\tau^{-}$ decay along the spin direction $\textbf{n}^{\pm}$ is related to the unpolarized decay rate by the following
relation:
\begin{equation}
\frac{d\Gamma (\textbf{n}^{\pm })}{ds}=\frac{1}{2}\left( \frac{d\Gamma }{%
ds}\right) \left[ 1+(P_{L}^{\pm }\textbf{e}_{L}^{\pm}+P_{N}^{\pm }\textbf{e}%
_{N}^{\pm }+P_{T}^{\pm }\textbf{e}_{T}^{\pm })\cdot \textbf{n}^{\pm }\right].
\label{polarized-decay}
\end{equation}%
Finally, by using the definitions (\ref{singlepol}), the expressions for the longitudinal $(P_{L})$, normal $(P_{N})$, and transverse $(P_{T})$
lepton polarizations can be written as
\begin{eqnarray}
P_{L}(s) &=&\frac{1}{\Delta}\bigg[\frac{4}{3} s v \left(m_H^2-s\right)^2
[2\mathcal{R}e(\mathcal{C}_1\mathcal{C}_2^*)+2\mathcal{R}e(\mathcal{C}_3\mathcal{C}_4^*)]-\frac{4m_{\tau}
\left(m_H^2-s\right)}{ s v^2}\{\tanh^{-1}(v)
\left(4 m_{\tau}^2-s v^2+s\right)\notag \\
\quad&&+s^2 v
\left(v^2-1\right)2\mathcal{R}e(\mathcal{C}_0\mathcal{C}_3^*)-\{2\left(s v^2+s\right) \tanh^{-1}(v)+sv(v^2-1)\}2\mathcal{R}e(\mathcal{C}_0\mathcal{C}_2^*)\}\bigg],\notag \\
   \label{pl1}
\end{eqnarray}
\begin{eqnarray}
P_{N}(s) &=&\frac{1}{\Delta}\bigg[-2\{\frac{4\pi m_{\tau}^2 m_H^2 \left(2 v^2+\sqrt{1-v^2}+1\right)}{\sqrt{s} v^2 \sqrt{1-v^2}}-\frac{\pi
   \sqrt{s} \left(\sqrt{1-v^2}+1\right) \left(4 m_\tau^2+m_H^2-s\right)}{v^2}+\frac{2 \sqrt{s} \left(m_H^2-2 s\right)}{v}\notag \\
&&+\frac{4 m_\tau^2 \sqrt{s} \left(2 \sqrt{1-v^2}+\pi  v\right)}{v
   \sqrt{1-v^2}}-\sqrt{s} \left(8
   m_\tau^2+(2+\pi ) m_H^2-(\pi -4) s\right)\}\mathcal{R}e(\mathcal{C}_0\mathcal{C}_2^*)\notag \\
&&+\{\frac{2\pi}{s^{3/2} v^2 \sqrt{1-v^2}}  \left(4 m_\tau^2 \left(m_H^2+s\right)+s \left(m_H^2-3 s\right)\right) \left(4 m_\tau^2-s
   \left(v^2+2 \sqrt{1-v^2}-1\right)\right)\} \mathcal{R}e(\mathcal{C}_0\mathcal{C}_3^*)\notag \\
&&+\pi  m_\tau \sqrt{s} \left(m_H^2-s\right)^2 [2 \mathcal{R}e(\mathcal{C}_1\mathcal{C}_3^*)]\bigg],
\label{pl2}
\end{eqnarray}
\begin{eqnarray}
P_{T}(s) &=& \frac{1}{\Delta}\bigg[\pi  v\sqrt{s}\left(\text{m}_H^2-s\right)\left\{ 2
\big[2\mathcal{I}m(\mathcal{C}_1^*\mathcal{C}_0)+2\mathcal{I}m(\mathcal{C}_4^*\mathcal{C}_0)\big]-m_\tau\left(\text{m}_H^2-s\right)
    \big[2\mathcal{I}m(\mathcal{C}_4^\ast\mathcal{C}_1)+2\mathcal{I}m(\mathcal{C}_2^\ast\mathcal{C}_3)\big]\right\}\bigg], \notag \\ \label{pl3}
\end{eqnarray}
where $\mathcal{C}_{0},\ldots,\mathcal{C}_{4}$ are given in Eq. (\ref{Cfunctions}) and the $\Delta$ used in the above equations is defined in Eq. (\ref{delta}).

To calculate the double-lepton-polarization asymmetries, we consider the
polarizations of both $\tau^{-}$ and $\tau^{+}$ simultaneously and introduce
the following spin projection operators for the $\tau^{-}$ and $\tau^{+}$,%
\begin{eqnarray}
\Lambda _{1} &=&\frac{1}{2}\left( 1+\gamma _{5}\slashed{S}_{i}^{-}\right),
\notag \\
\Lambda _{2} &=&\frac{1}{2}\left( 1+\gamma _{5}\slashed{S}_{i}^{+}\right),
\label{DLPV}
\end{eqnarray}%
where $i=L,T\,\ $and $N$ again designate the longitudinal, transverse and
normal lepton polarizations index, respectively. In the rest frame of the
$\tau^{-}\tau^{+}$ one can define the following set of orthogonal vectors $%
S^{\alpha }$:%
\begin{eqnarray}
S_{L}^{-\alpha } &=&(0,\textbf{e}_{L}^{-})=\left( 0,\frac{\textbf{p}_{1}}{\left\vert
\textbf{p}_{1}\right\vert }\right) ,  \notag \\
S_{N}^{-\alpha } &=&(0,\textbf{e}_{N}^{-})=\left( 0,\frac{\textbf{k}\times \textbf{p}_{1}%
}{\left\vert \textbf{k}\times \textbf{p}_{1}\right\vert }\right) ,  \notag \\
S_{T}^{-\alpha } &=&(0,\textbf{e}_{T}^{-})=\left( 0,\textbf{e}_{N}^{-}\times \textbf{e}%
_{L}^{-}\right) ,  \label{DLvectors} \\
S_{L}^{+\alpha } &=&(0,\textbf{e}_{L}^{+})=\left( 0,\frac{\textbf{p}_{2}}{\left\vert
\textbf{p}_{2}\right\vert }\right) ,  \notag \\
S_{N}^{+\alpha } &=&(0,\textbf{e}_{N}^{+})=\left( 0,\frac{\textbf{k}\times \textbf{p}_{2}%
}{\left\vert \textbf{k}\times \textbf{p}_{2}\right\vert }\right) ,  \notag \\
S_{T}^{+\alpha } &=&(0,\textbf{e}_{T}^{+})=\left( 0,\textbf{e}_{N}^{+}\times \textbf{e}%
_{L}^{+}\right) .  \notag
\end{eqnarray}%
Just like the single lepton polarizations, through Lorentz transformations we
can boost the longitudinal component in the CM frame of $\tau ^{-}\tau ^{+}$
as%
\begin{eqnarray}
\left( S_{L}^{-\alpha }\right) _{CM} &=&\left( \frac{|\textbf{p}_{1}|}{m_{l}},%
\frac{E\textbf{p}_{1}}{m_{l}\left\vert \textbf{p}_{-}\right\vert }\right),  \notag
\\
\left( S_{L}^{+\alpha }\right) _{CM} &=&\left( \frac{|\textbf{p}_{2}|}{m_{l}},-%
\frac{E\textbf{p}_{2}}{m_{l}\left\vert \textbf{p}_{+}\right\vert }\right).
\label{DLboosted}
\end{eqnarray}%
The normal and transverse components remain the same under Lorentz boost. We
now define the double lepton polarization asymmetries as%
\begin{equation}
P_{ij}(s)=\frac{\left( \frac{d\Gamma }{ds}\left( \textbf{S}_{i}^{-},%
\textbf{S}_{j}^{+}\right) -\frac{d\Gamma }{ds}\left( -\textbf{S}_{i}^{-},\textbf{S%
}_{j}^{+}\right) \right) -\left( \frac{d\Gamma }{ds}\left( \textbf{S}%
_{i}^{-},-\textbf{S}_{j}^{+}\right) -\frac{d\Gamma }{ds}\left( -\textbf{S}%
_{i}^{-},-\textbf{S}_{j}^{+}\right) \right) }{\left( \frac{d\Gamma }{ds}%
\left( \textbf{S}_{i}^{-},\textbf{S}_{j}^{+}\right) -\frac{d\Gamma }{ds}\left(
-\textbf{S}_{i}^{-},\textbf{S}_{j}^{+}\right) \right) +\left( \frac{d\Gamma }{%
ds}\left( \textbf{S}_{i}^{-},-\textbf{S}_{j}^{+}\right) -\frac{d\Gamma }{ds%
}\left( -\textbf{S}_{i}^{-},-\textbf{S}_{j}^{+}\right) \right) },
\label{DLdefinition}
\end{equation}%
where the subscripts $i$ and $j$ correspond to the $\tau^{-}$ and $\tau^{+}$
polarizations indices, respectively. Using these definitions the various double lepton polarization
asymmetries as a function of $s$ can be written as
\begin{eqnarray}
P_{LL}(s) &=&\frac{1}{\Delta}\bigg[\frac{32}{s^2 v^3 \left(m_H^2-s\right)^2} \{\tanh^{-1}(v) (-64 m_{\tau}^6 s+8 m_{\tau}^4 (m_H^4+7 s^2)-4
m_{\tau}^2 s(m_H^4-m_H^2 s+5 s^2)\notag \\
&&+s^2(m_H^4+2 m_H^2 s (v^2-1)+s^2 (3-2v^2)))-s v (16 m_{\tau}^4 s-2 m_{\tau}^2 (m_H^4+3 s^2)\notag \\
&&+s(m_H^4-m_H^2s+s^2))\}\left\vert\mathcal{C}_0\right\vert^2+\frac{(m_H^2-s)^2}{3m_{\tau}^2}\{(8 m_{\tau}^4+s^2
(v^2-1))(\left\vert\mathcal{C}_1\right\vert^2+\left\vert\mathcal{C}_3\right\vert^2)\notag
\\
&&+s^2 v^2(v^2-1)
(\left\vert\mathcal{C}_4\right\vert^2+\left\vert\mathcal{C}_2\right\vert^2)\}-\frac{8}{m_{\tau} s v^3}\{m_{\tau}^2 s v(2s v^2-2(m_H^2-s)-v^3(m_H^2-s))
   \notag \\
&&+ \tanh ^{-1}(v)(64 m_{\tau}^6-16 m_{\tau}^4 s-8 m_{\tau}^2s^2-s^2 (v^2-1)(2s+v^3(m_H^2-s)))\}(2\mathcal{R}_e(\mathcal{C}_0\mathcal{C}_1^*))\notag
\\
&&+ \frac{2}{m_{\tau} s} \{(m_H^2-s)(s (4 m_{\tau}^2+s v^{5/2}-s
v^3)-\tanh ^{-1}(v) ((4 m_{\tau}^2+s)^2-s^2
v^3))(2\mathcal{R}_e(\mathcal{C}_0\mathcal{C}_4^*))\}\bigg],\notag
   \\
 \label{pll1}
\end{eqnarray}
\begin{eqnarray}
P_{NN}(s)&=&\frac{1}{\Delta}\bigg[\frac{32}{v(m_H^2-s)^2}\{(2(m_H^2-s) (2 m_{\tau}^2-s)+s^2 v^2 (v^2+1)) \tanh ^{-1}(v)+s v (m_H^2+s(v^2-1)\}\left\vert\mathcal{C}_0\right\vert^2\notag \\
 &&+\frac{2}{3} s v^2(m_H^2-s)^2\{(\left\vert\mathcal{C}_1\right\vert^2+\left\vert\mathcal{C}_3\right\vert^2)-(\left\vert\mathcal{C}_4\right\vert^2+\left\vert\mathcal{C}_2\right\vert^2)\}-\frac{8
 m_{\tau}}{v} \{8 m_{\tau}^2 \tanh ^{-1}(v)+s (\pi -2 v)\}(2\mathcal{R}_e(\mathcal{C}_0\mathcal{C}_1^*))\bigg],\notag \\
\label{pll5}
\end{eqnarray}
\begin{eqnarray}
P_{TT}(s)&=&\frac{1}{\Delta}\bigg[-\frac{32}{s^2 v^3 (m_H^2-s)^2} (16
m_{\tau}^4 s+2 m_{\tau}^2 (m_H^4-4 m_H^2 s-s^2)+m_H^2
s^2)
(s v-s(v^2+1)\tanh ^{-1}(v))\left\vert\mathcal{C}_0\right\vert^2\notag \\
&&+\frac{16 m}{s v^3}(2 m_{\tau}^2 \tanh ^{-1}(v) (-m_H^2+2 s
v^2+s)-s^2
v^3)(2\mathcal{R}e(\mathcal{C}_0\mathcal{C}_1^*))\notag \\
&&+\frac{2}{3} s (m_H^2-s)^2 \big((2-v^2) (\left\vert\mathcal{C}_1\right\vert^2+\left\vert\mathcal{C}_3\right\vert^2)-v^2
(\left\vert\mathcal{C}_4\right\vert^2+\left\vert\mathcal{C}_2\right\vert^2)\big)\bigg],\label{pll9}
\end{eqnarray}
\begin{eqnarray}
P_{LN}(s)&=&\frac{1}{\Delta}\bigg[\frac{\pi }{2 v^2} \sqrt{s}(m_H^2-s)
\{-4 \sqrt{1-v^2}(\frac{2 m_{\tau}^2}{s}+v^2)-\frac{8m_{\tau}^2}{s}+2
   v^2+2\} ((2\mathcal{I}m(\mathcal{C}_0\mathcal{C}_2^\ast))\notag \\
   &+&2 \pi  \sqrt{s}(\sqrt{1-v^2}-1) (m_H^2-s) (2\mathcal{I}m(\mathcal{C}_0\mathcal{C}_3^*))\bigg]=-P_{NL}(s),\label{pll2}
\end{eqnarray}
\begin{eqnarray}
P_{LT}(s)&=&\frac{1}{\Delta}\bigg[-\frac{16 \pi  m_{\tau}}{\sqrt{s} \sqrt{\frac{1}{v^2}-1}}\left\vert\mathcal{C}_0\right\vert^2\notag \\
   &&+\frac{1}{2} \pi  m_{\tau} \sqrt{s} v \left(m_H^2-s\right)^2 (2\mathcal{R}e(\mathcal{C}_1\mathcal{C}_4^*)+2\mathcal{R}e(\mathcal{C}_2\mathcal{C}_3^*))-\frac{2 \pi  \sqrt{s} \left(\sqrt{1-v^2}-1\right) \left(m_H^2-s\right)}{v}(2\mathcal{R}e(\mathcal{C}_0\mathcal{C}_1^*))\notag \\
   &&+\frac{\left(m_H^2-s\right)}{2 s^{3/2} v^2}\{\pi  \left\{-16 m_{\tau}^4+8 m_{\tau}^2 s+s^2
   \left(v^4+3\right)\right\}-8 \pi  m_{\tau} s  \sqrt{2 m_{\tau}^2+s
   v^2}\}(2\mathcal{R}e(\mathcal{C}_0\mathcal{C}_4^*))\bigg]\-P_{TL},\label{pll3}
\end{eqnarray}
\begin{eqnarray}
P_{NT}(s)&=&\frac {8 m_{\tau}} {\Delta v^2}  \bigg[\left (m_H^2 -
s \right)\left(\left (v^2 - 1 \right) \tanh ^{-1} (v) +
v \right) (2\mathcal {I} m (\mathcal {C} _ 0\mathcal {C}_3^\ast)) + \left (4 m_{\tau}^2\tanh^{-1} (v) -
s v \right) (2\mathcal {I} m (\mathcal {C} _ 0\mathcal {C}_2^\ast))\bigg]\notag \\
&=&-P_{TN},\label{pll6}
\end{eqnarray}
with $\mathcal{C}_{0},\ldots,\mathcal{C}_{4}$ and $\Delta$ used in above equations are defined in Eq. (\ref{Cfunctions}) and Eq. (\ref{delta}), respectively.
\begin{table}[ht]
\centering \caption{Default values of input parameters used in the
calculations}
\begin{tabular}{c}
\hline\hline
$m_{H}=125$ GeV, $m_{t}=172$ GeV, $m_{\mu}=0.106$ GeV,\\
$m_e=0.51\times10^{-3}$ GeV, $m_{\tau}=1.77$ GeV, $\alpha(M_{Z})^{-1}=128$,
\\  $m_Z=91.18$ GeV, $\Gamma_Z=2.48$ GeV.\\
\hline\hline
\end{tabular}\label{input}
\end{table}
\section{Numerical Analysis}
In the previous section we have presented the expressions of the single and
double lepton polarization asymmetries in the SM
for $H\to\gamma\tau^+\tau^-$ decay by considering both
tree and loop diagrams. To proceed with the numerical analysis of these physical observables, the numerical values of different
input parameters in the SM are listed in Table 1.


\subsection{Single Lepton Polarization Asymmetries}

Similar to the case of leptons forward-backward asymmetries $A_{FB}$ calculated in ref. \cite{AFB1}, one can see from Eqs. (\ref{pl1}, \ref{pl2}, \ref{pl3}) that the single
lepton polarization asymmetries are not separately dependent on the
tree level diagrams. However, they depend on the interference of the contributions from different loop diagrams (c.f. Figure \ref{loopdiag}) as well as on the interference of tree and loop diagrams. Moreover, for the
longitudinal polarization asymmetry $P_L$ the contributions generated
from the interference between tree and loop diagrams $\mathcal C_{0} \mathcal C_{3}^*$ and $\mathcal C_{0} \mathcal C_{2}^*$ are
$m_{\tau}$ suppressed.

\begin{figure}
\centering
\includegraphics[scale=0.6]{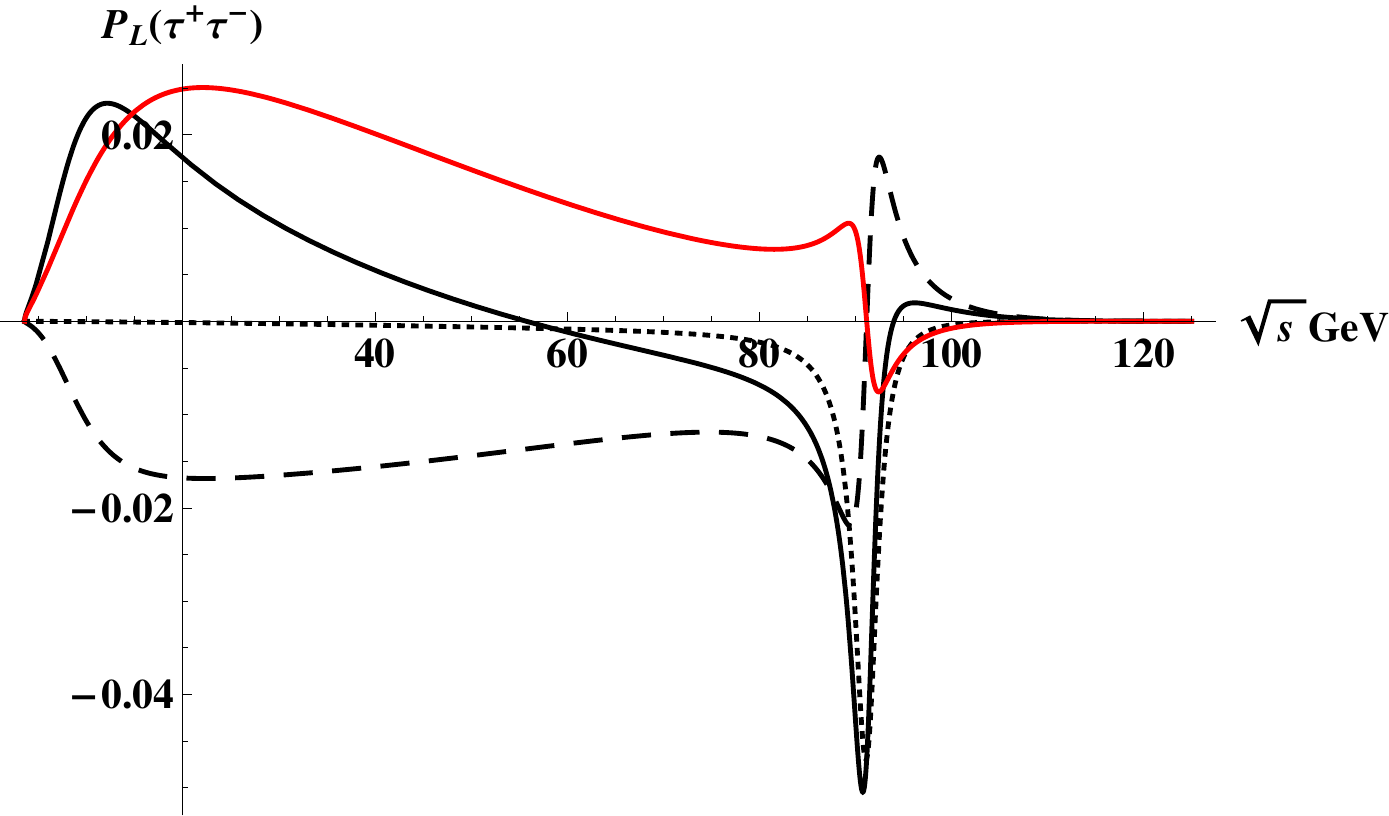}
\caption{The longitudinal lepton polarization asymmetry $P_L(s)$ in
$H\to\gamma \tau^+ \tau^-$ as a function of the momentum
transfer $\sqrt{s}$. The solid line correspond to the total contribution,
the dashed line correspond to the contribution through the
interference between the tree diagram and $Z^\ast$ pole diagrams and
the dotted line corresponds to the contribution from the
$Z^\ast-Z^\ast$ interference while the red line corresponds to the
contribution from the $\gamma^\ast-Z^\ast$ interference.}\label{Pltaus}
\end{figure}

\begin{figure}
\centering
\includegraphics[scale=0.6]{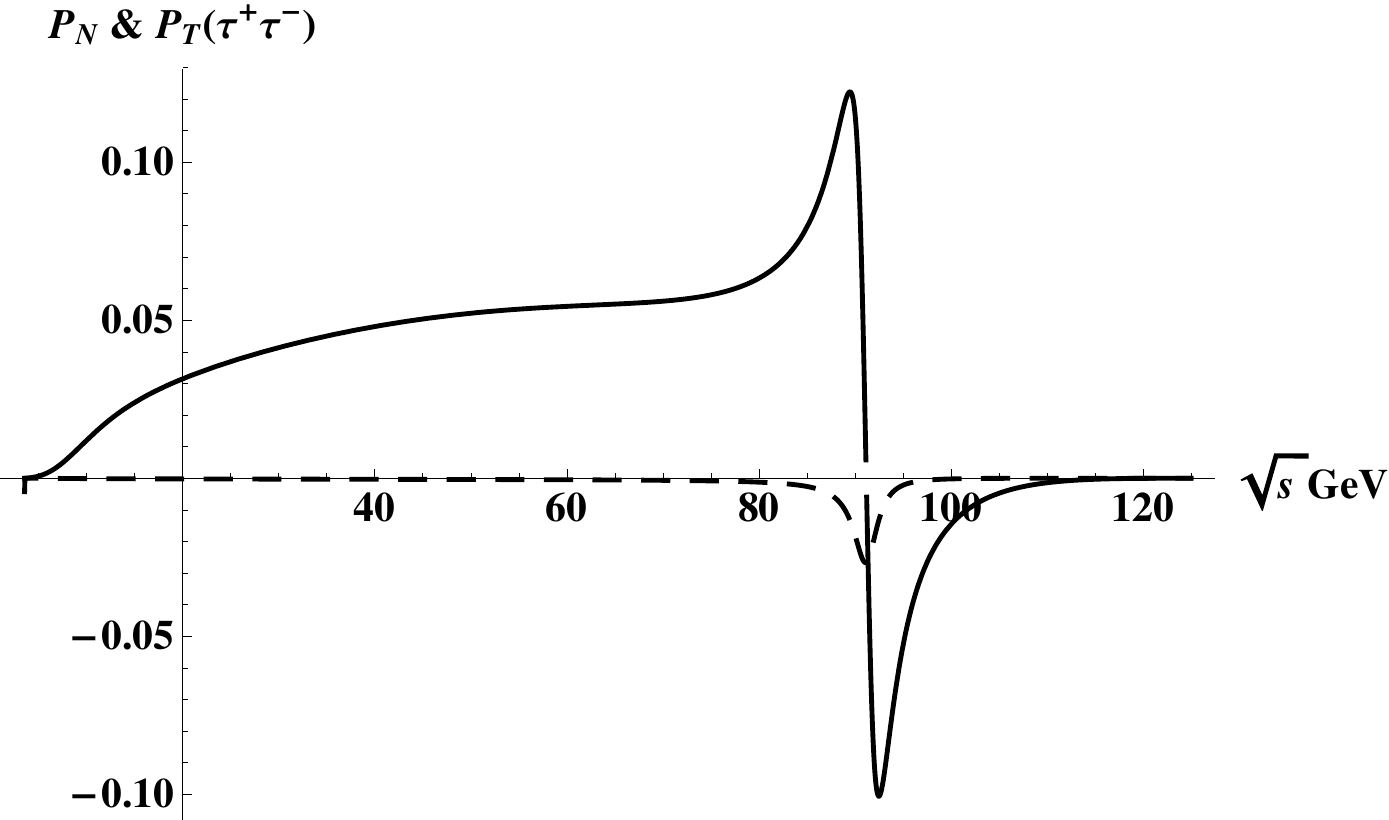}
\caption{The Normal lepton polarization asymmetry $P_N(s)$ in
$H\to\gamma\tau^+\tau^-$ as a function of the momentum
transfer $\sqrt{s}$. The solid line corresponds to the $P_{N}$ and the dashed line is for $P_{T}$.}\label{PNfig}
\end{figure}

\begin{table*}
\begin{tabular}{c|c|c|c|c|c|cc}
\hline\hline
$s_{min}-s_{max}(\text{GeV}^2)$&$10^2-30^2$&$30^2-50^2$&$50^2-70^2$&$70^2-90^2$&$90^2-110^2$&full
Phase space
\\ \hline
 \ \  $H\to\gamma\tau^+\tau^-$ \ \  & \ \  3.3\% \ \  & \ \  4.8\% \ \  & \ \ 5.4\% \ \  & \ \  7.9\% \ \  & \ \  -1.3\% \ \  & \ \  0.02\% \ \    \\
\hline\hline
\end{tabular}
\caption{The average Normal polarization asymmetries $\langle
P_{N}\rangle$ for some cuts on $s$ in $H\to\gamma \tau^+\tau^-$ decays .} \label{wc table}
\end{table*}

The longitudinal lepton polarization $P_L$ as a function of $\sqrt{s}$ in
$H \to \gamma\tau^+\tau^-$ is drawn in Fig. \ref{Pltaus}. It is expected from Eq. (\ref{pl1}) that the
major contribution is coming through the interference of different
loop diagrams. If we recall Eq. (\ref{Cfunctions}) for the definitions of
$\mathcal{C}_1-\mathcal{C}_4$ then there are two types of
contributions, namely, $\gamma^*-Z^*$ interference (referred as the
overlap between the $\gamma^*$ pole amplitude and the $Z^*$ pole
amplitude) and the second one is $Z^*-Z^*$ interference (i.e. the
interference among the different amplitudes of $Z^*$ pole diagrams).
Similar
to the $A_{FB}$ \cite{AFB1} the longitudinal polarization asymmetry also
has zero crossing because of the change of sign of  $\gamma^*-Z^*$ interference at $s=m_Z^2$ which is due to the sign change of the real part of $P_Z$ and it is evident from Fig. \ref{Pltaus}. It can also be seen from Figure \ref{Pltaus} that throughout the allowed
kinematical region the value of $P_L$ in $H\to\gamma\tau^+\tau^-$
is of the order of $10^{-2}$ and to measure such a small number at the current colliders is not an easy task.

In contrast to the longitudinal lepton
polarization asymmetry $(P_L)$ and the leptons forward-backward
asymmetry $(A_{FB})$, the contributions from the loop diagrams (c.f. contributions from the $Z^*-Z^*$ and $\gamma^*-Z^*$ interference )
are $m_{\tau}$ suppressed in the normal lepton polarization asymmetry $(P_N)$ and hence can be safely ignored.
The normal lepton polarization ($P_N$), where the contribution is coming from the
interference between the tree and $Z^*$ pole diagrams is displayed with solid line in Figure \ref{PNfig}.


Just like the leptons forward-backward asymmetry  $(A_{FB})$ calculated in ref. \cite{AFB1}, the magnitude of
transverse lepton polarization asymmetry $(P_{T})$ is very small in almost all the kinematical
region and the dahsed line in Figure \ref{PNfig} demonstrate this fact. By looking at the Eq. (\ref{pl3}) it is even more evident that it is proportional
to the imaginary part of the contributions from different interference diagrams which are too small, hence we ignored its detailed
analysis over here.

Furthermore, we have also calculated the average value of normal polarization asymmetry $(P_{N})$
in different bins of $s$ and listed it in Tables 2 for
$H\to\gamma\tau^+\tau^-$. It is clear from
the Table 2 that by scanning full phase space altogether this
asymmetry might not be measurable quantities at LHC, but by making an analysis
in different bins of $s$ the
average value of normal lepton polarization asymmetry
may be measurable at the LHC.

\subsection{Double Lepton Polarization Asymmetries}

Now we discuss the double lepton polarization asymmetries $P_{ij}$, where the indices $i,j$ can be
$L$, $T$ and $N$.
In Section 3, we have derived the expressions of different double lepton polarization asymmetries (c.f. Eqs. (\ref{pll1}-\ref{pll9})),
where one can immediately notice that in contrast to the above discussed single lepton polarization
asymmetries and the forward-backward asymmetries, discussed in length in ref. \cite{AFB1}, some of the
$P_{ij}$'s are explicit functions of the tree diagrams in addition to their interference with loop diagrams.
\begin{table*}
\begin{tabular}{c|c|c|c|c|c|cc}
\hline\hline
$s_{min}-s_{max}(\text{GeV}^2)$&$10^2-30^2$&$30^2-50^2$&$50^2-70^2$&$70^2-90^2$&$90^2-110^2$&full
Phase space
\\ \hline
 \ \  $H\to\gamma\tau^+\tau^-$ \ \  & \ \ 86.7\% \ \  & \ \  80.8\% \ \  & \ \ 84.7\% \ \  & \ \  84.6\% \ \  & \ \  93.8\% \ \  & \ \  99.6\% \ \    \\
\hline\hline
\end{tabular}
\caption{The average polarization asymmetries $\langle
P_{LL}\rangle$ for some cuts on $s$ in $H\to\gamma \tau^+\tau^-$ decays .}
\end{table*}

\begin{figure}
\centering
\includegraphics[scale=0.6]{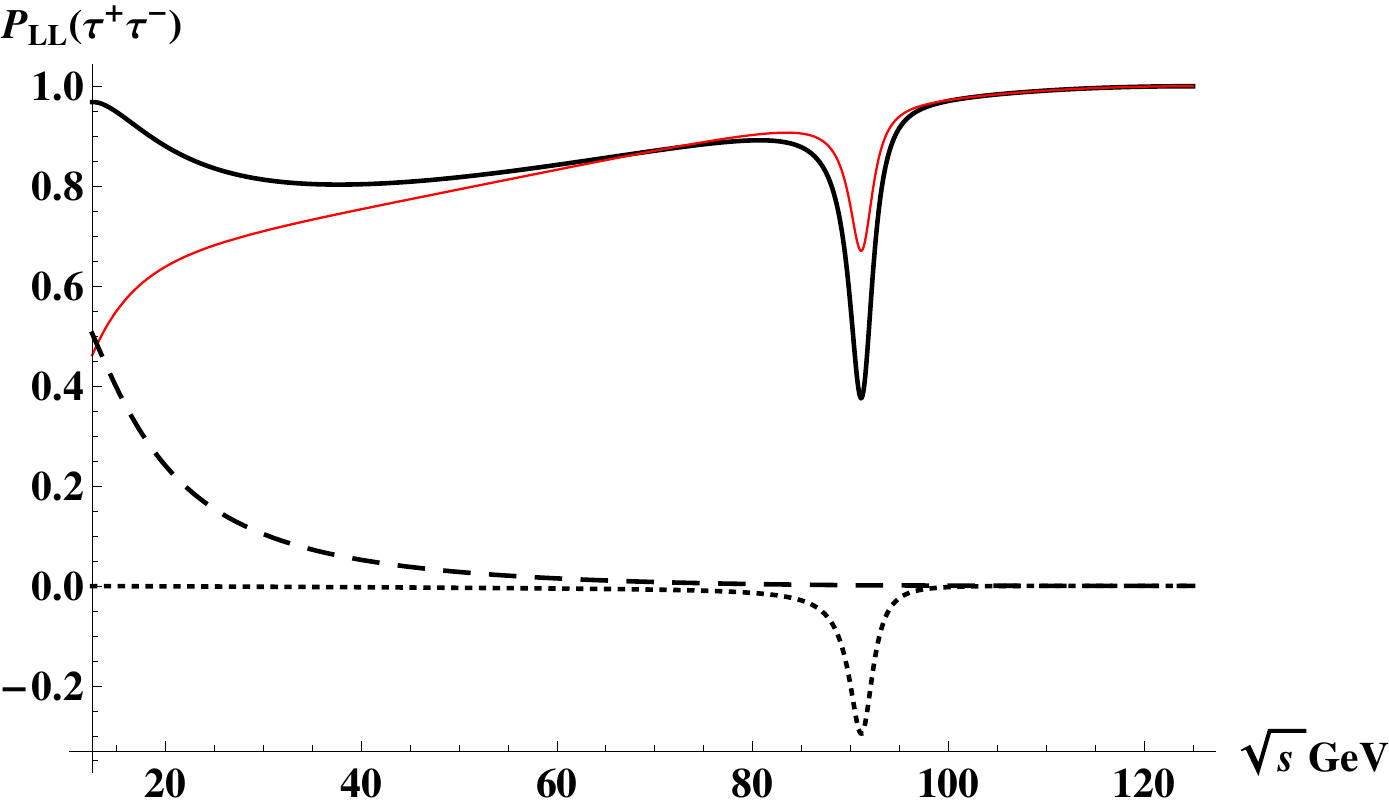}
\caption{$P_{LL}(s)$ in $H\to\gamma \tau^+ \tau^-$ as a function
of the momentum transfer $\sqrt{s}$. The solid line correspond to the total
contribution, the red line correspond to the contribution coming
from only tree diagrams and the dotted line corresponds to the
contribution from the $Z^\ast-Z^\ast$ interference while the dashed
line corresponds to the contribution from the interference
of tree diagrams and $\gamma^*$
diagrams. The other contributions are too small to be shown
here.}\label{PLLTaus}
\end{figure}

The double longitudinal lepton polarization asymmetry ($P_{LL}$) in
$H\to\gamma\tau^+\tau^-$ as a function of $\sqrt{s}$ is drawn in Fig. \ref{PLLTaus}. Fig. \ref{PLLTaus} shows that in the $H\to\gamma\tau^+\tau^-$ decay the
major contribution is coming from the tree diagram which is
denoted by the red line in this figure. The average value of $P_{LL}$ in various bins of $s$ is given in Table 3 where we can see that this asymmetry
has a significantly large value in the whole phase space range. From experimental point of view, it has already been mentioned that
the reconstruction of the polarization is possible only if the particle is unstable and due to the decay of $\tau$ lepton to a final state charged hadron its spin direction is inferred from the decay products distribution \cite{Sug1, Sug2} at the LHC. Therefore, the scanning of double longitudinal lepton
polarization asymmetry at the LHC will help us to dig out some intrinsic properties of the Higgs particle and
its dynamics.

\begin{table*}
\begin{tabular}{c|c|c|c|c|c|cc}
\hline\hline
$s_{min}-s_{max}(\text{GeV}^2)$&$10^2-30^2$&$30^2-50^2$&$50^2-70^2$&$70^2-90^2$&$90^2-110^2$&full
Phase space
\\ \hline
\ \  $H\to\gamma\tau^+\tau^-$ \ \  & \ \ -7\% \ \  & \ \  18\% \ \  & \ \ 41.7\% \ \  & \ \  64.6\% \ \  & \ \  88.7\% \ \  & \ \  99\% \ \    \\
\hline\hline
\end{tabular}
\caption{The average polarization asymmetries $\langle
P_{TT}\rangle$ for some cuts on $s$ in $H\to\gamma \tau^+\tau^-$ decay.}
\end{table*}

\begin{figure}
\centering
\includegraphics[scale=0.6]{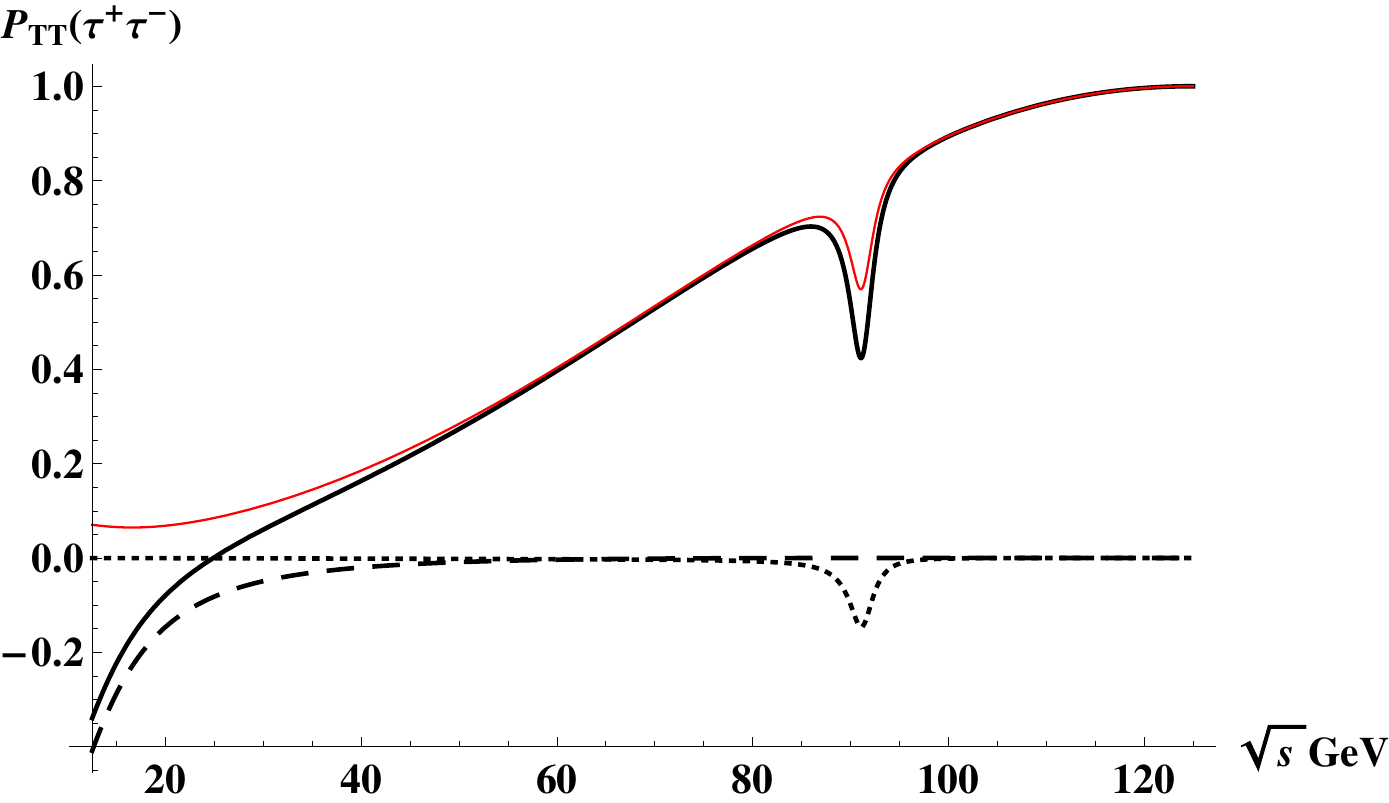}
\caption{$P_{TT}(s)$ in $H\to\gamma \tau^+ \tau^-$ as a function of the momentum transfer $\sqrt{s}$. The legends are same as in Fig. \ref{PLLTaus}.}\label{PTTtaus}
\end{figure}

Fig. \ref{PTTtaus} display the behavior of $P_{TT}$ as a function of $\sqrt{s}$ for $H\to\gamma\tau^+\tau^-$ decay where the major contribution is coming from the tree diagram and it is denoted by the red line in Fig. \ref{PTTtaus}.
In Fig. \ref{PTTtaus} one can see that in the region $90$GeV$\leq
\sqrt{s}\leq \sqrt s_{max}$ the major contribution is coming only from the tree
diagrams and it can be seen that the average value of $\langle P_{TT} \rangle$ enhanced upto
$88\%$ in this region (c.f. Table 5) which is in a measurable ball park of the LHC.

\begin{table*}
\begin{tabular}{c|c|c|c|c|c|cc}
\hline\hline
$s_{min}-s_{max}(\text{GeV}^2)$&$10^2-30^2$&$30^2-50^2$&$50^2-70^2$&$70^2-90^2$&$90^2-110^2$&full
Phase space
\\ \hline
\ \  $H\to\gamma\tau^+\tau^-$ \ \  & \ \ -6.2\% \ \  & \ \  -20.2\% \ \  & \ \ -42.6\% \ \  & \ \  -68.7\% \ \  & \ \  -90.8\% \ \  & \ \  -99.1\% \ \    \\
\hline\hline
\end{tabular}
\caption{The average polarization asymmetries $\langle
P_{NN}\rangle$ for some cuts on $s$ in $H\to\gamma \tau^+\tau^-$ decays.}
\end{table*}

\begin{figure}
\centering
\includegraphics[scale=0.6]{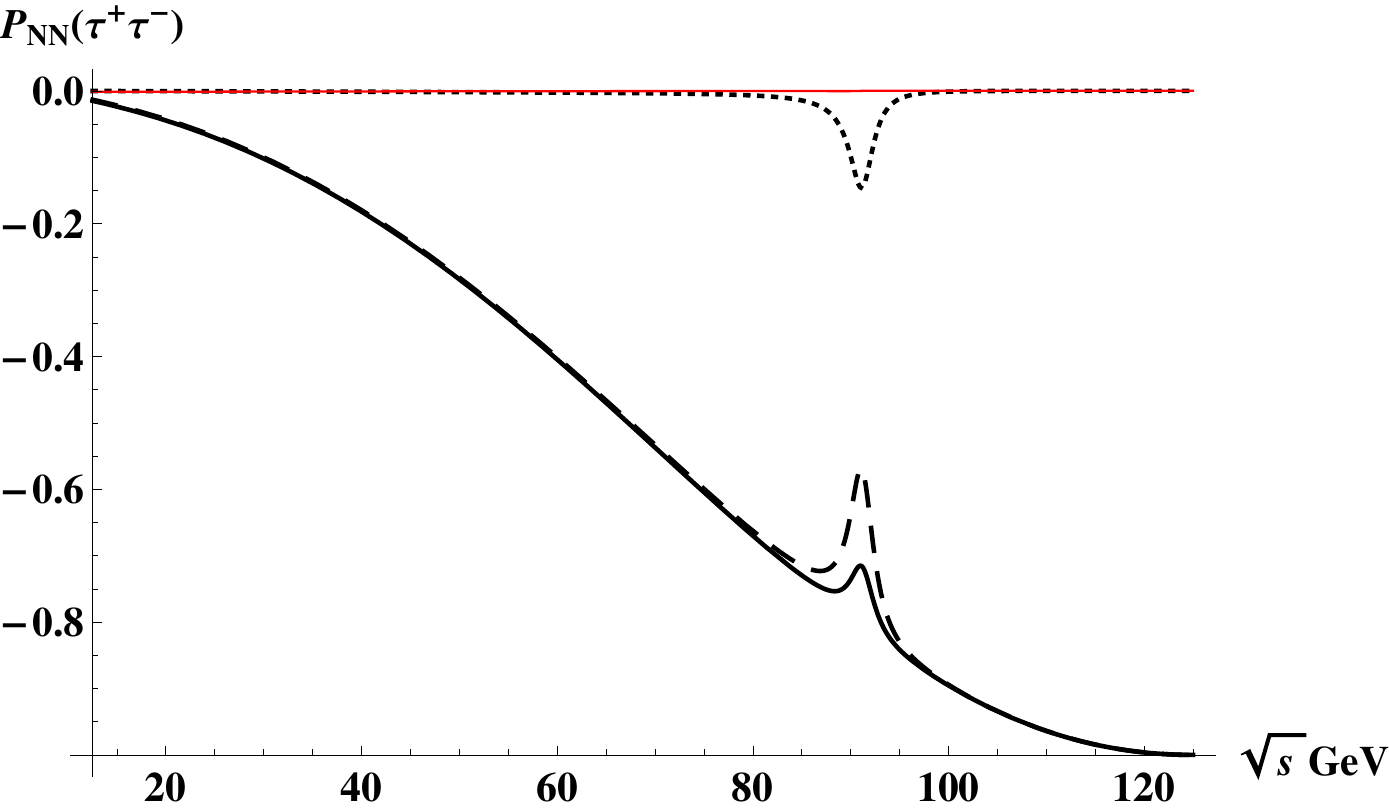}
\caption{$P_{NN}(s)$ in $H\to\gamma \tau^+ \tau^-$ as a function
of the momentum transfer $\sqrt{s}$. The solid line correspond to the total
contribution, the dotted line corresponds the $Z^*-Z^*$ interference
contribution and the red line corresponds to the contribution from
$Z^*-\gamma^*$ interference while the dashed line corresponds to the
contribution from the tree diagrams.  The other contributions are
too small to be shown here.}\label{PNNtaus}
\end{figure}

Similarly, $P_{NN}$ in $H\to\gamma\tau^+\tau^-$ as a function of $\sqrt{s}$ is drawn in Fig.
\ref{PNNtaus}. From Fig. \ref{PNNtaus} we can notice that
in $H\to\gamma\tau^+\tau^-$, the terms contributing to $P_{NN}$ are coming from the
tree diagrams. In order to make the analysis more clear the values of
average $P_{NN}$ in various bins of $s$ are given in the Table 6. We can see that in the region
$50$GeV$<\sqrt{s}<110$GeV, the average values of double normal lepton polarization asymmetry enhanced
up to $91 \%$ for $H \to \gamma \tau^{+} \tau^{-}$ decay. Therefore, it will be interesting to scan the $P_{NN}$
in different bins of $s$ to clear the smog from the Higgs decays.

\begin{table*}
\begin{tabular}{c|c|c|c|c|c|cc}
\hline\hline
$s_{min}-s_{max}(\text{GeV}^2)$&$10^2-30^2$&$30^2-50^2$&$50^2-70^2$&$70^2-90^2$&$90^2-110^2$&full
Phase space
\\ \hline
 \ \  $H\to\gamma\tau^+\tau^-$ \ \  & \ \ -43.6\% \ \  & \ \  -28.1\% \ \  & \ \ -15.7\% \ \  & \ \  -6.5\% \ \  & \ \  -1.7\% \ \  & \ \  -4.2\% \ \    \\
\hline\hline
\end{tabular}
\caption{The average polarization asymmetries $\langle
P_{LT}\rangle$ for some cuts on $s$ in $H\to\gamma \tau^+\tau^-$ decay.}
\end{table*}

\begin{figure}
\centering
\includegraphics[scale=0.6]{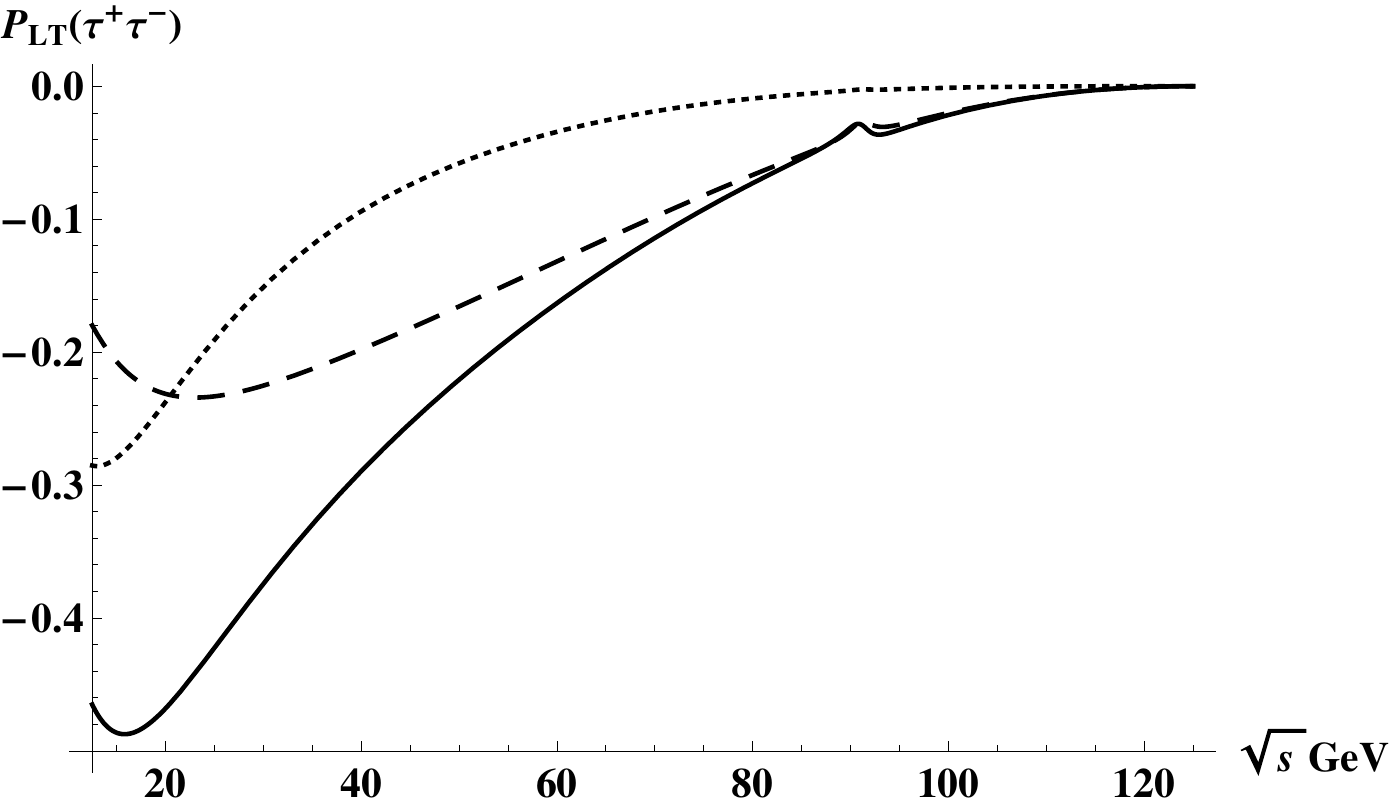}
\caption{$P_{LT}(s)$ in $H\to\gamma \tau^+ \tau^-$ as a function of
the momentum transfer $\sqrt{s}$. The solid line correspond to
the total contribution, the red line corresponds to the contribution
coming from the interference between tree diagrams and $Z^*$ pole
diagrams and the dotted line corresponds to the contribution from
the interference between tree diagrams and the $\gamma^*$ diagrams
while the dashed line corresponds to the contribution from the tree
diagrams only. The other contributions are negligible.}\label{PLTtaus}
\end{figure}

In $H\to\gamma\tau^+\tau^-$ the $P_{LT}$ as a function of $\sqrt{s}$ is displayed in Fig. \ref{PLTtaus}, where one can see that the
major contribution is through the tree diagrams, represented by the dashed
line in Fig. \ref{PLTtaus}. The contributions from the other diagrams come out to be negligible
to show in the graphs. Finally, the average values of $P_{LT}$ is listed in Table 7 in various bins of the $s$. Remarkably, in contrast to the lepton forward-backward
asymmetry $A_{FB}$ \cite{AFB1}, like other polarization asymmetries the value of $P_{LT}$ for $H \to \gamma \tau^+ \tau^-$ is significantly large in different regions of $s$, and we hope that
these can be measured at the LHC.

\begin{figure}
\centering
\includegraphics[scale=0.6]{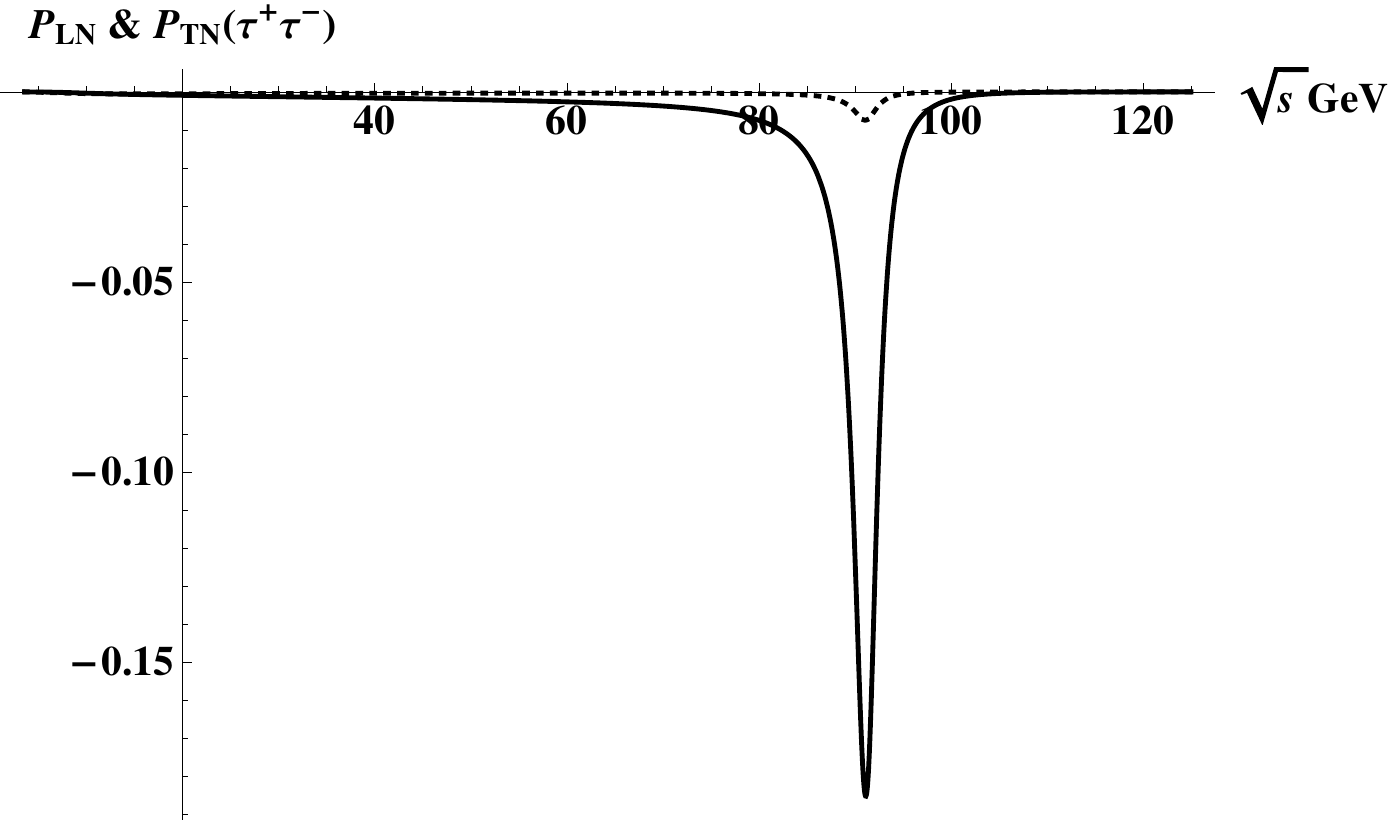}
\caption{$P_{LN}(s)$ and $P_{TN}$ as a function  of the momentum
transfer $\sqrt{s}$ in $H \to \gamma \tau^{+}\tau^{-}$ decay. The solid line corresponds to the $P_{LN}$ and
 the dashed line refers to the $P_{TN}$.}\label{PLNtauon}
\end{figure}

In Fig. \ref{PLNtauon} we have plotted $P_{LN}$ and $P_{TN}$ as a function of $\sqrt{s}$ in $H \to \gamma \tau^{+}\tau^{-}$ decay, where the solid
and dashed lines correspond to the $P_{LN}$ and $P_{TN}$, respectively.
From Eqs. (\ref{pll2}) and (\ref{pll6}), it is clear that these two
asymmetries are proportional to the imaginary part of the contributions arising from different diagrams, encoded in $\mathcal{C}'s$ and hence
their value is expected to be small and it is clear from Fig. \ref{PLNtauon}.
In case of $P_{LN}$, the
contribution is coming from the interference between tree and $Z^*$
pole diagrams and for $P_{TN}$ it is coming from the
interference between tree and $\gamma^*$ diagrams, while the other
contributions are small enough to plot at the present scale.

\section{Conclusion}

In this work we have calculated the single and double lepton
polarization asymmetries in $H\to\gamma\tau^+\tau^-$ decay.
 To calculate these asymmetries we consider both the
tree and loop level diagrams. First, we have derived the expressions
of these asymmetries and then explicitly evaluated the SM
contributions of these diagrams to $P_i$ and $P_{ij}$, where $i,j$ can be $L$, $N$ or $T$.  We have found
that both type of diagrams are significant in the calculation of the
SM values of the above mentioned asymmetries. The analysis shows that the
separate contribution of the tree level diagrams for the single
lepton polarization asymmetries is absent, just like in
the forward-backward asymmetries $A_{FB}$ \cite{AFB1}, which however is present in the
the double lepton polarization asymmetries. Particularly,
in $P_{LL}$ and $P_{TT}$ for $H\to\gamma\tau^+\tau^-$ the contributions from the tree level diagrams
are the dominant one.

As one of the major
motivation of this study is to see either the magnitude of the
asymmetries other than the forward-backward asymmetries is large
enough to be measured or not. Regarding this, interestingly, the numerical calculations
performed here show that most of the single and double
lepton polarization asymmetries comes out to be an order of magnitude larger
than the forward-backward asymmetries $A_{FB}$ calculated in ref.\cite{AFB1}.

To highlight, by analyzing the different single and double lepton polarization asymmetries
in $H \to \gamma \tau^{+}\tau^{-}$ decay,
we found that the average values of these asymmetries in some kinematical region are sufficiently
large (c.f. Tables 4-7) than the forward-backward asymmetries $A_{FB}$
\cite{AFB1} which can be measured at the current colliders.
To keep an eye on the features of the $A_{FB}$, $P_i$ and $P_{ij}$
we would comparatively assume that the later polarization
asymmetries are good observable in $H\to\gamma\tau^+\tau^-$. Therefore, the measurement of different single and double lepton polarization asymmetries for $H \to \gamma \tau^+ \tau^-$ decay will definitely help out to clear some smog from the intrinsic properties of Higgs boson and
its dynamics.

\appendix
\section{Appendix}
The function $P_Z $ is defined as
\begin{eqnarray}
P_{Z}=\frac{1}{sin\theta_{W}
cos\theta_{W}}\frac{1}{s-m_{Z}^2+im_{Z}\Gamma_{Z}},
\end{eqnarray}
where $s=q^2 = (p_1 + p_2)^2$ is square of momentum transfer to the lepton pair, $\theta_{W}$ is the weak mixing angle and
$\Gamma_{Z}$ is the decay width of $Z$ boson.
The functions $\Pi_{\alpha \gamma Z}$,  $\Pi_{s\gamma Z}$ and
$\Pi_{\gamma \gamma}$ are induced from the $Z^*$ and $\gamma^*$ pole
diagrams given in Fig. 2 and these can be summarized as follows:
\begin{eqnarray}
\Pi_{a\gamma
Z}=\frac{\alpha^2}{m_{W}sin\theta_{W}}\frac{N_{c}Q_{f}T_{f}}{sin{\theta_{W}}cos\theta_{W}}A_{f_2}(\tau_{f},\lambda_{f}),
\end{eqnarray}
\begin{eqnarray}
\Pi_{s\gamma
Z}=\frac{\alpha^2}{m_{W}sin\theta_{W}}\bigg[-cot\theta_{W}A_{W}(\tau_{W},\lambda_{W})\notag\\
-2N_{c}Q_{f}
\frac{T_{f}-2Q_{f}sin^2\theta_{W}}{sin\theta_{W}cos\theta_{W}}A_{f_{1}}(\tau_{f},\lambda_{f})\bigg],
\end{eqnarray}
\begin{eqnarray}
\Pi_{\gamma \gamma
}=\frac{\alpha^2}{m_{W}sin\theta_{W}}\bigg[-A_{W}(\tau_{W},\lambda_{W})-4N_{c}Q_{f}^2
A_{f_{1}(\tau_{f},\lambda_{f})}\bigg].
\end{eqnarray}

Here, the functions $A_{W}$, $A_{f_1}$ and $A_{f_2}$ denote the contributions from the $W$ boson and fermion loops,
with $\tau = \frac{4m^2_{i}}{m^2_{H}}$, $\lambda_{i} = \frac{4m^2_{i}}{q^2} (i = f, W)$, $m_{f}$ is the fermion mass, $N_{c}$
the color multiplication, $Q_{f}$ is fermion charge, $\tau_{f}$ is the third component of weak isospin of fermion $f$ inside
the loop. Expressions of these loop functions can be summarized as
\begin{eqnarray}
A_{f_1}(\tau, \lambda)=I_{1}(\tau, \lambda)-I_{2}(\tau, \lambda),
\end{eqnarray}
\begin{eqnarray}
A_{f_2}(\tau, \lambda)=\frac{\tau
\lambda}{\lambda-\tau}[2g(\tau)-2g(\lambda)+f(\tau)-f(\lambda)],
\end{eqnarray}
\begin{eqnarray}
A_{W}(\tau, \lambda)=\bigg[
\left(1+\frac{2}{\tau}\right)\left(\frac{4}{\lambda}-1\right)-\left(5+\frac{2}{\tau}\right)\bigg]I_{1}(\tau,
\lambda)+16\left(1-\frac{1}{\lambda}\right)I_{2}(\tau, \lambda)
\end{eqnarray}
with
\begin{eqnarray}
I_{1}(\tau,
\lambda)=\frac{\tau\lambda}{2(\tau-\lambda)}+\frac{\tau^2
\lambda^2}{2(\tau-\lambda)^2}[f(\tau)-f(\lambda)]+\frac{\tau^2
\lambda}{(\tau-\lambda)^2}[g(\tau)-g(\lambda)],
\end{eqnarray}
\begin{eqnarray}
I_{2}(\tau, \lambda)=-\frac{\tau
\lambda}{2(\tau-\lambda)}[f(\tau)-f(\lambda)],
\end{eqnarray}
and
\begin{eqnarray}
f(\tau)&=&\begin{cases} arcsin^2(\tau^{(-\frac{1}{2})}) & \text{for
} \tau\geq1
\\
-\frac{1}{4}\bigg[\ln\frac{1+\sqrt{1-\tau}}{1-\sqrt{1-\tau}}-i\pi\bigg]^{2}{
} & \text{for }  \tau <1,
\end{cases}
\end{eqnarray}

\begin{eqnarray}
g(\tau)&=&\begin{cases} \sqrt {\tau -1 }
arcsin(\tau^{(-\frac{1}{2})}) & \text{for }\tau\geq1
\\
\frac{\sqrt
{1-\tau}}{2}\bigg[\ln\frac{1+\sqrt{1-\tau}}{1-\sqrt{1-\tau}}-i\pi\bigg]
& \text{for } \tau <1.
\end{cases}
\end{eqnarray}


\begin{thebibliography}{99}
\bibitem{4} G. Aad \textit{et al.} [ATLAS Collaboration], \emph{Observation of a new particle in the
search for the Standard Model Higgs boson with the ATLAS detector at
the LHC, Phys. Lett. B}\textbf{716} (2012) 1; [arXiv:1207.7214[hep-ex]].

\bibitem{5} S. Chatrchyan \textit{et al.} [CMS Collaboration], \emph{Observation of a new boson at a mass
of 125 GeV with the CMS experiment at the LHC, Phys. Lett. B}\textbf{716}
(2012) 30; [arXiv:1207.7235[hep-ex]]; \emph{Observation of a new boson with
mass near 125 GeV in PP collisions at ps=7 TeV and 8 TeV},
arXiv:1303.4571[hep-ex].

\bibitem{N1} F.~J.~Petriello, \emph{Kaluza-Klein effects on Higgs physics in universal extra dimensions, JHEP} {\bf 0205}, 003 (2002) [arXiv: hep-ph/0204067].
\bibitem{N1a} T.~Han, H.~E.~Logan, B.~McElrath and L.~-T.~Wang, \emph{Loop induced decays of the little Higgs: $H\to gg, \gamma \gamma$  Phys.\ Lett.\ B} {\bf 563}, 191 (2003) [Erratum-ibid.\ B {\bf 603}, 257 (2004)] [arXiv: hep-ph/0302188].
\bibitem{N1b} G.~Cacciapaglia, A.~Deandrea and J.~Llodra-Perez,  \emph{$H \to \gamma \gamma$ beyond the Standard Model, JHEP} {\bf 0906}, 054 (2009) [arXiv:0901.0927 [hep-ph]].
\bibitem{N1c} J.~Cao, Z.~Heng, T.~Liu and J.~M.~Yang, \emph{Di-photon Higgs signal at the LHC: A Comparative study for different supersymmetric models, Phys.\ Lett.\ B} {\bf 703}, 462 (2011) [arXiv:1103.0631 [hep-ph]].

\bibitem{N2}  S.~Heinemeyer, O.~Stal and G.~Weiglein, \emph{Interpreting the LHC Higgs Search Results in the MSSM, Phys.\ Lett.\ B} {\bf 710}, 201 (2012) [arXiv:1112.3026 [hep-ph]].
\bibitem{N2a} P.~M.~Ferreira, R.~Santos, M.~Sher and J.~P.~Silva, \emph{Implications of the LHC two-photon signal for two-Higgs-doublet models, Phys.\ Rev.\ D} {\bf 85}, 077703 (2012) [arXiv:1112.3277 [hep-ph]].
\bibitem{N2b} K.~Cheung and T.~-C.~Yuan, \emph{Could the excess seen at 124-126 GeV be due to the Randall-Sundrum Radion?, Phys.\ Rev.\ Lett.}  {\bf 108}, 141602 (2012) [arXiv:1112.4146 [hep-ph]].
\bibitem{N2c} N.~Chen and H.~-J.~He, \emph{LHC Signatures of Two-Higgs-Doublets with Fourth Family, JHEP} {\bf 1204}, 062 (2012) [arXiv:1202.3072 [hep-ph]].
\bibitem{N2d} S.~Dawson and E.~Furlan, \emph{A Higgs Conundrum with Vector Fermions,  Phys.\ Rev.\ D} {\bf 86}, 015021 (2012) [arXiv:1205.4733 [hep-ph]].
\bibitem{N2e} M.~Carena, I.~Low and C.~E.~M.~Wagner, \emph{Implications of a Modified Higgs to Diphoton Decay Width, JHEP} {\bf 1208}, 060 (2012) [arXiv:1206.1082 [hep-ph]].
\bibitem{N2f} J.~Chang, K.~Cheung, P.~-Y.~Tseng and T.~-C.~Yuan, \emph{Distinguishing Various Models of the 125 GeV Boson in Vector Boson Fusion, JHEP} {\bf 1212}, 058 (2012) [arXiv:1206.5853 [hep-ph]].
\bibitem{N3} H.~An, T.~Liu and L.~-T.~Wang, \emph{125 GeV Higgs Boson, Enhanced Di-photon Rate, and Gauged \textit{$U(1)_{PQ}$-Extended} MSSM, Phys.\ Rev.\ D} {\bf 86}, 075030 (2012) [arXiv:1207.2473 [hep-ph]].
\bibitem{N3a} T.~Abe, N.~Chen and H.~-J.~He, \emph{LHC Higgs Signatures from Extended Electroweak Gauge Symmetry, JHEP} {\bf 1301}, 082 (2013) [arXiv:1207.4103 [hep-ph]].
\bibitem{N3b} A.~Joglekar, P.~Schwaller and C.~E.~M.~Wagner, \emph{Dark Matter and Enhanced Higgs to Di-photon Rate from Vector-like Leptons, JHEP} {\bf 1212}, 064 (2012) [arXiv:1207.4235 [hep-ph]].
\bibitem{N3c} L.~G.~Almeida, E.~Bertuzzo, P.~A.~N.~Machado and R.~Z.~Funchal, \emph{Does $H \to \gamma \gamma$ Taste like vanilla New Physics?, JHEP} {\bf 1211}, 085 (2012) [arXiv:1207.5254 [hep-ph]].
\bibitem{N3d} A.~Delgado, G.~Nardini and M.~Quiros, \emph{Large diphoton Higgs rates from supersymmetric triplets, Phys.\ Rev.\ D} {\bf 86}, 115010 (2012) [arXiv:1207.6596 [hep-ph]].
\bibitem{N3e} M.~Hashimoto and V.~A.~Miransky, \emph{Enhanced diphoton Higgs decay rate and isospin symmetric Higgs boson, Phys.\ Rev.\ D} {\bf 86}, 095018 (2012) [arXiv:1208.1305 [hep-ph]].
\bibitem{N3f} T.~Kitahara, \emph{Vacuum Stability Constraints on the Enhancement of the h --> gamma gamma rate in the MSSM, JHEP} {\bf 1211}, 021 (2012) [arXiv:1208.4792 [hep-ph]].
\bibitem{N3g} S.~Chang, S.~K.~Kang, J.~-P.~Lee, K.~Y.~Lee, S.~C.~Park and J.~Song, \emph{Comprehensive study of two Higgs doublet model in light of the new boson with mass around 125 GeV, JHEP} {\bf 1305}, 075 (2013) [arXiv:1210.3439 [hep-ph]].
\bibitem{N31h} G.~Moreau, \emph{Constraining extra-fermion(s) from the Higgs boson data, Phys.\ Rev.\ D} {\bf 87}, 015027 (2013)
  [arXiv:1210.3977 [hep-ph]].
\bibitem{N3h} M.~Chala,  \emph{$h \rightarrow \gamma\gamma$ excess and Dark Matter from Composite Higgs Models, JHEP} {\bf 1301}, 122 (2013) [arXiv:1210.6208 [hep-ph]].
\bibitem{N3i} S.~Dawson, E.~Furlan and I.~Lewis, \emph{Unravelling an extended quark sector through multiple Higgs production?,
  Phys.\ Rev.\ D} {\bf 87}, 014007 (2013) [arXiv:1210.6663 [hep-ph]].
\bibitem{N3j} K.~Choi, S.~H.~Im, K.~S.~Jeong and M.~Yamaguchi, \emph{Higgs mixing and diphoton rate enhancement in NMSSM models, JHEP} {\bf 1302}, 090 (2013) [arXiv:1211.0875 [hep-ph]].
\bibitem{N3k} C.~Han, N.~Liu, L.~Wu, J.~M.~Yang and Y.~Zhang, \emph{Two-Higgs-doublet model with a color-triplet scalar: a joint explanation for top quark forward-backward asymmetry and Higgs decay to diphoton} [arXiv:1212.6728 [hep-ph]].
\bibitem{N3l} W.~-Z.~Feng and P.~Nath, \emph{Higgs diphoton rate and mass enhancement with vector-like leptons and the scale of supersymmetry, Phys.\ Rev.\ D} {\bf 87}, 075018 (2013) [arXiv:1303.0289 [hep-ph]].
\bibitem{N3m} T.~Kitahara and T.~Yoshinaga, \emph{Stau with Large Mass Difference and Enhancement of the Higgs to Diphoton Decay Rate in the MSSM, JHEP} {\bf 1305}, 035 (2013) [arXiv:1303.0461 [hep-ph]].

\bibitem{6} A. Abbasabadi, D. Browser-Chao, D.A. Dicus and W.W. Repko, \emph{Radiative Higgs boson decays $H\rightarrow \gamma f f$ , Phys. Rev. D} \textbf{55} (1997) 5647; [arXiv: hep-ph/9611209].
\bibitem{8} A. Abbasabadi and W.W. Repko, Phys. Rev. D 62 (2000) 054025; [arXiv:hep-ph/0004147].
\bibitem{9} Ana Firan and Ryszard Stroynowski, \emph{Internal conversions in Higgs decays to two photons, Phys. Rev. D} \textbf{76} (2007) 057301,
arXiv:0704.3987[hep-ph].
\bibitem{10} L.B. Chen, C.F. Qiao and R.L. Zhu, \emph{Reconstructing the 125 GeV SM Higgs boson,} arXiv:1211.6058[hep-ph] .
\bibitem{11} D.A. Dicus and W.W. Repko, \emph{Calculation of the decay}, arXiv:1302.2159[hep-ph].

\bibitem{AFB1} Y.~Sun, H.~-R.~Chang and D.~-N.~Gao, \emph{Higgs decays to $\gamma l^+ l^-$ in the standard model,JHEP} {\bf 1305}, 061 (2013)
  [arXiv:1303.2230 [hep-ph]].

 \bibitem{Sug1} S. Chatrchyan \textit{et al.} (CMS Collaboration), \emph{Measurement of the Polarization of $W$ Bosons with Large Transverse Momenta in $W+jets$ Events at the LHC, Phys. Rev. Lett.} {\bf 107}, 021802 (2011) [arXiv:1104.382[hep-ex]].

\bibitem{Sug2} G. Aad \textit{et al.} (ATLAS Collaboration), \emph {Measurement of the $W$ boson polarization in top quark decays with the ATLAS detector, JHEP} {\bf 1206}, 088 (2012) [arXiv:1205.2484[hep-ex]].

\bibitem{Polarization1} F.~Kruger and L.~M.~Sehgal, \emph{Lepton Polarization in the Decays $B\to X_s\mu^+\mu^-$ and $B\to X_s\tau^+\tau^-$,'' Phys.\ Lett.\  B} {\bf 380}, 199 (1996) [arXiv:hep-ph/9603237].


\bibitem{Sug3} G. Aad \textit{et al.} (ATLAS Collaboration), \emph{Measurement of tau polarization in $W\rightarrow \tau\nu$ decays with the ATLAS detector in pp collisions at $\sqrt{s} = 7 TeV$, Eur. Phys. J. C} {\bf 72}, 2062 (2012) [arXiv:1204.6720 [hep-ex]].

\bibitem{Polarization2}  S.~Fukae, C.~S.~Kim and T.~Yoshikawa, \emph{A systematic analysis of the lepton polarization asymmetries in the  rare B decay, $B \to X_s \tau^+ \tau^-$,'' Phys.\ Rev.\  D} {\bf 61}, 074015 (2000) [arXiv:hep-ph/9908229].


\end{thebibliography}
\end{document}